\documentstyle[prb,aps,twocolumn]{revtex}

\input{epsf}

\begin{document}

\twocolumn[\hsize\textwidth\columnwidth\hsize\csname
@twocolumnfalse\endcsname

\draft

\title{Symmetry Constraints and Variational Principles in Diffusion
Quantum Monte Carlo Calculations of Excited-State Energies}

\author{W. M. C. Foulkes} 
\address{Blackett Laboratory, Imperial College of Science, Technology
and Medicine, Prince Consort Road, London SW7 2BZ, England}

\author{Randolph Q. Hood\cite{address} and R. J. Needs}
\address{Cavendish Laboratory, Madingley Road, Cambridge CB3 0HE,
England}

\date{\today}

\maketitle

\begin{abstract}
\begin{quote}
Fixed-node diffusion Monte Carlo (DMC) is a stochastic algorithm for
finding the lowest energy many-fermion wave function with the same
nodal surface as a chosen trial function.  It has proved itself among
the most accurate methods available for calculating many-electron
ground states, and is one of the few approaches that can be applied to
systems large enough to act as realistic models of solids.  In
attempts to use fixed-node DMC for excited-state calculations, it has
often been assumed that the DMC energy must be greater than or equal
to the energy of the lowest exact eigenfunction with the same symmetry
as the trial function.  We show that this assumption is not justified
unless the trial function transforms according to a one-dimensional
irreducible representation of the symmetry group of the Hamiltonian.
If the trial function transforms according to a multi-dimensional
irreducible representation, corresponding to a degenerate energy
level, the DMC energy may lie below the energy of the lowest
eigenstate of that symmetry.  Weaker variational bounds may then be
obtained by choosing trial functions transforming according to
one-dimensional irreducible representations of subgroups of the full
symmetry group.
\end{quote}
\end{abstract}

\narrowtext
]

\section{Introduction}
\label{sec:Introduction}

Quantum Monte Carlo (QMC) methods are powerful and general tools for
calculating the ground-state electronic properties of atoms,
molecules, and solids.  Since the computational cost increases only as
the cube of the number of particles, it is possible to study systems
containing hundreds of electrons subject to periodic boundary
conditions.  This is enough to model real condensed matter with
surprising precision, as shown by the accuracy of 0.1 eV per atom or
better achieved in QMC calculations of the cohesive energies of
solids.  By comparison, the errors in local density functional
calculations of cohesive energies are often of the order of 1 eV per
atom.

The two most widely used QMC methods are variational Monte Carlo (VMC)
and diffusion Monte Carlo (DMC).~\cite{dmc,hammond} In VMC a trial
many-electron wave function is chosen and expectation values are
evaluated using Monte Carlo integration, which is more efficient than
grid-based quadrature methods for high-dimensional integrals.  Most
VMC simulations of solids use trial wave functions containing a number
of adjustable parameters, the values of which are determined by
minimizing the energy or its variance.

DMC is a stochastic method for evolving a solution of the
imaginary-time Schr\"{o}dinger equation.  The imaginary-time evolution
gradually enhances the ground-state component of the solution relative
to the excited-state components, but the algorithm does not maintain
the fermionic symmetry of the starting state.  The solution therefore
converges towards the overall ground state, which is bosonic.  This
difficulty is known as the sign problem.

Although several exact solutions to the sign problem have been
proposed, none has the statistical efficiency required to study the
large systems of interest to condensed matter physicists.  Most DMC
simulations therefore use the approximate fixed-node
method,~\cite{anderson} which is numerically stable and often very
accurate.  The details will be described in Sec.~\ref{sec:DMC}, but
the basic idea is quite simple.  A real trial many-electron wave
function is chosen and used to define a trial nodal surface, which is
the surface on which the trial function is zero and across which it
changes sign.  In a three-dimensional system containing $N$ electrons,
the trial wave function is a function of $3N$ variables, and the trial
nodal surface is $3N-1$ dimensional in general.  The fixed-node DMC
algorithm maintains the nodal surface of the trial wave function, so
enforcing the fermionic symmetry and producing the lowest energy
many-electron wave function consistent with that nodal surface.

Although VMC and DMC are principally ground-state methods, they can
also provide some information about excited states.  In particular,
they can be used to study the lowest energy state of each distinct
symmetry.  In VMC this is done by choosing a trial wave function which
possesses the required symmetry for all values of the variational
parameters.  The energy obtained after optimizing the trial function
is therefore greater than or equal to the eigenvalue of the lowest
energy eigenstate of that symmetry.  A similar technique is also used
in DMC, although this is much harder to justify.  The problem is that
the DMC trial function is only used to define the trial nodal surface,
which may not be sufficient to fix the symmetry of the state produced
by the stochastic DMC algorithm.  In any case, practical tests have
shown that this approach often gives excellent results.  Examples are
the study of excitations of the hydrogen molecule by Grimes {\it et
al.},~\cite{grimes} and calculations of excitation energies in
diamond~\cite{mitas2,mitas3} and silicon.~\cite{sibands}

If the trial function used in an excited-state DMC simulation has no
definite symmetry, the only certainty is that the DMC energy must be
greater than or equal to the many-electron ground-state
energy.~\cite{hammond} In cases when the DMC trial function does have
a definite symmetry, however, it is normally assumed that the
fixed-node DMC solution has the same symmetry as the trial function,
and hence that the DMC energy is greater than or equal to the
eigenvalue of the lowest energy eigenstate of that symmetry.  This
symmetry-constrained variational principle is widely accepted, but we
show by constructing a specific example that it is not always correct:
the fixed-node DMC solution need not have the same symmetry as the
trial function; and the fixed-node DMC energy may be lower than the
energy of the lowest exact eigenstate of that symmetry.

The symmetry-constrained DMC variational principle is guaranteed to
hold only when the trial function transforms according to a
one-dimensional irreducible representation of the symmetry group of
the Hamiltonian.  The corresponding eigenstate is then non-degenerate,
or has only accidental degeneracies.  If the trial function transforms
according to a multi-dimensional irreducible representation,
corresponding to a degenerate energy level, the DMC energy may lie
below the energy of the lowest eigenstate of that symmetry.  In such
cases a weaker variational principle may be obtained by choosing a
trial function that transforms according to a one-dimensional
irreducible representation of a subgroup of the full symmetry group.
The DMC energy is then greater than or equal to the eigenvalue of the
lowest exact eigenstate with that subgroup symmetry.  This provides a
strict variational lower bound for the DMC energy, but one that
usually lies below the energy of the degenerate eigenstate of
interest.

As an example, consider the case of a crystalline solid.  Any trial
function with a definite crystal momentum ${\bf k}$ satisfies the
many-electron version of Bloch's theorem and so transforms according
to a one-dimensional irreducible representation of the translation
group, which is a subgroup of the full symmetry group.  The weaker
variational principle therefore guarantees that the DMC energy must be
greater than or equal to the energy of the lowest exact eigenstate
with crystal momentum ${\bf k}$.  Unfortunately, most Bloch states are
complex and so cannot be used as fixed-node DMC trial functions.  Real
linear combinations of Bloch functions and their complex conjugates
can be used instead, but in most cases these do not transform
according to one-dimensional irreducible representations and do not
lead to useful variational principles.  This is illustrated in
Sec.~\ref{sec:sepexample}, where we show that the DMC energy obtained
using such a trial function may lie below the energy of the lowest
eigenstate with crystal momentum ${\bf k}$.

The weaker variational principle is useful, but relies on a very
careful choice of trial functions and cannot explain all the past
successes of the fixed-node DMC method for excited states.  The real
explanation of these successes, we believe, is that although the DMC
algorithm does not always preserve the symmetry of the trial function,
the imposed nodal surface acts as such a strong restriction that the
DMC solution cannot stray ``too far'' from that symmetry.  The
calculated energy is therefore close to the variational value that
would have been obtained if the symmetry had been preserved.  In cases
when the excited state of interest satisfies the strong variational
principle, the errors in the ground and excited state energies are
guaranteed to have the same sign and tend to cancel, so improving the
accuracy of the calculated energy difference.

The rest of this paper is organized as follows: Sec.~\ref{sec:DMC}
contains brief explanations of the DMC method and the fixed-node
approximation for ground states; Sec.~\ref{sec:GenVar} shows that no
general symmetry-constrained variational principle exists;
Sec.~\ref{sec:Lowest} shows that a variational theorem holds for the
lowest energy state of each symmetry provided that the trial function
transforms according to a one-dimensional irreducible representation
of the group of spatial transformations of the Hamiltonian;
Sec.~\ref{sec:d>1} introduces a weaker variational principle which may
give energy bounds even when the trial function transforms according
to an irreducible representation of dimension greater than one;
Sec.~\ref{sec:tiling} shows that a generalization of the tiling
theorem~\cite{tiling} holds in every case when we can prove that the
DMC energy obeys a variational bound; Secs.~\ref{sec:tbexample} and
\ref{sec:sepexample} give examples which demonstrate the absence of a
general symmetry-constrained variational principle and illustrate the
application of the weaker variational principle; and
Sec.~\ref{sec:Conclusions} summarizes and concludes.

\section{Fixed-Node Diffusion Monte Carlo for Ground States}
\label{sec:DMC}

In this section we summarize the known results concerning the
application of the fixed-node DMC method to ground
states.~\cite{dmc,hammond} The aim is to evaluate expectation values
with an antisymmetric wave function $\Phi(X)$, where $X \equiv (x_1,
x_2, \ldots, x_N)$ lists the coordinates of all $N$ electrons, and
$x_i = ({\bf r}_i, s_{zi})$ specifies the position and spin projection
of electron $i$.  We choose wave functions with a fixed total $S_z =
\sum_{i=1}^{N} s_{zi}$.  The expectation value of a spin-independent
symmetric operator $\hat{A}({\bf R})$ is given by

\begin{equation}
\label{eq:Aspin}
\langle A \rangle = \frac {\sum_S \int \Phi^*(X) \hat{A}({\bf
R}) \Phi(X) \, d{\bf R}} {\sum_S \int \Phi^*(X)
\Phi(X) \, d{\bf R}} \;,
\end{equation}

\noindent
where ${\bf R} \equiv ({\bf r}_1,{\bf r}_2,\ldots,{\bf r}_N)$.  For
each spin configuration, $S$, the antisymmetric wave function $\Phi(X)
= \Phi(x_1,x_2,\ldots,x_N)$ may be replaced by a version with permuted
arguments, $\Phi(x_{i_1}, x_{i_2}, \ldots, x_{i_N})$, where the
permutation is chosen such that the first $N_{\uparrow}$ arguments are
spin-up and the last $N_{\downarrow}=N-N_{\uparrow}$ are spin-down.
Since ${\bf R}$ is a dummy variable, we can relabel $({\bf
r}_{i_1},{\bf r}_{i_2},\ldots, {\bf r}_{i_N})$ as $({\bf r}_1,{\bf
r}_2,\ldots,{\bf r}_N)$, after which the sums over spin configurations
can be removed.  The expectation value may then be written as

\begin{equation}
\label{eq:Anospin}
\langle A \rangle = \frac {\int \tilde{\Phi}^*({\bf R}) \hat{A}({\bf
R}) \tilde{\Phi}({\bf R}) \, d{\bf R}} {\int \tilde{\Phi}^*({\bf R})
\tilde{\Phi}({\bf R}) \, d{\bf R}} \;,
\end{equation}

\noindent where

\begin{eqnarray}
\lefteqn { \tilde{\Phi}({\bf r}_1,{\bf r}_2,\ldots,{\bf r}_N) \; \equiv
} \nonumber & & \\
& \;\;\; & \Phi({\bf r}_1,\uparrow; {\bf r}_2,\uparrow;\ldots;
{\bf r}_{N_{\uparrow}},\uparrow; {\bf r}_{N_{\uparrow}+1},\downarrow; 
\ldots;{\bf r}_{N},\downarrow) \;.
\end{eqnarray}

\noindent
The function $\tilde{\Phi}({\bf r}_1,\ldots,{\bf r}_{N})$ is
antisymmetric under interchange of any two of the up-spin arguments
${\bf r}_1,\ldots,{\bf r}_{N_{\uparrow}}$ or any two of the down-spin
arguments ${\bf r}_{N_{\uparrow}+1},\ldots,{\bf
r}_{N_{\uparrow}+N_{\downarrow}}$, but has no definite symmetry under
interchange of up- and down-spin arguments.  It still obeys the
Schr\"{o}dinger equation, but the up- and down-spin electrons are now
treated as distinguishable, allowing us to avoid explicit reference to
the spin variables.  For simplicity, all wave functions in the
following discussion will be chosen to be of this type.

DMC is a method for solving the imaginary-time Schr\"{o}dinger
equation,

\begin{equation}
\label{eq:its}
\left ( - \frac{1}{2}\nabla_{\bf R}^2 + V({\bf R}) - E_{\rm S} \right
) \Psi({\bf R},\tau) = - \frac{\partial}{\partial \tau} \Psi({\bf
R},\tau) \;,
\end{equation}

\noindent 
where $\nabla_{\bf R}^2$ is shorthand for $\sum_{i=1}^N \nabla_i^2$.
The potential $V({\bf R})$ includes the electron-electron interactions
as well as the external potential terms, and is assumed to be a local
function of ${\bf R}$.  It has Coulomb singularities whenever two
charged particles approach each other, but is finite everywhere else.
The constant energy shift $E_{\rm S}$ has been introduced to set a
convenient zero of energy as explained below.  The variable $\tau$
(which is real) is usually called the imaginary time, but we will
often abbreviate ``imaginary time'' to ``time'' in what follows.

If the starting state $\Psi({\bf R},\tau=0)$ is written as a linear
combination of energy eigenfunctions, 

\begin{equation}
\label{eq:lincomb}
\Psi({\bf R},\tau=0) = \sum_i
c_i \Psi_i({\bf R}) \; ,
\end{equation}

\noindent
the large $\tau$ limit of the solution of Eq.~(\ref{eq:its}) takes the
form,

\begin{equation}
\Psi({\bf R},\tau \rightarrow \infty)  =
c_l \Psi_l({\bf R}) e^{-(E_l - E_{\rm S}) \tau} 
\; ,
\end{equation}

\noindent
where $\Psi_l({\bf R})$ is the lowest energy eigenfunction appearing
in Eq.~(\ref{eq:lincomb}) and $E_l$ is the corresponding eigenvalue.
In applications to many-electron systems, $\Psi({\bf R},\tau=0)$ is
antisymmetric and $\Psi_l({\bf R})$ is usually the many-electron
ground state $\Psi_0({\bf R})$.  We will assume that the Hamiltonian
$\hat{H}$ is real (possesses time-reversal symmetry), in which case
$\Psi_l({\bf R})$ may also be chosen real and may be obtained by
following the imaginary-time evolution of a real function $\Psi({\bf
R},\tau)$.  If the Hamiltonian is complex (does not possess
time-reversal symmetry), as when there is an applied magnetic field,
the fixed-node DMC method discussed in this paper does not apply and
it is necessary to use the fixed-phase method of Ortiz {\it et
al.}~\cite{ortiz}

DMC solves Eq.~(\ref{eq:its}) using a stochastic algorithm, the
efficiency of which is much improved by an importance sampling
transformation.~\cite{reynolds} A real trial wave function $\Phi_{\rm
T}({\bf R})$ is chosen and Eq.~(\ref{eq:its}) is recast in terms of
the product

\begin{equation}
\label{eq:f}
f({\bf R},\tau) = \Psi({\bf R},\tau) \Phi_{\rm T}({\bf R})
\end{equation}

\noindent
which is also real.  After some straightforward algebra, $f({\bf
R},\tau)$ is found to satisfy the equation,

\begin{eqnarray}
- \frac{\partial}{\partial \tau}f({\bf R},\tau) & = & 
- \frac{1}{2}\nabla_{\bf R}^{2}f({\bf R},\tau) \; + \; \nabla_{\bf R} \cdot
\left[{\bf F}({\bf R})f({\bf R},\tau)\right] \nonumber \\
&& \; + \; \left(E_{\rm L}({\bf R}) - E_{\rm S}\right)f({\bf R},\tau) \; ,
\label{eq:fits}
\end{eqnarray}

\noindent 
where $E_{\rm L}({\bf R}) \equiv \Phi_{\rm T}^{-1}({\bf R}) \hat{H}
\Phi_{\rm T}({\bf R})$ is known as the local energy and ${\bf F}({\bf
R}) \equiv \Phi_{\rm T}^{-1}({\bf R}) \nabla_{\bf R}\Phi_{\rm T}({\bf
R})$ as the quantum force.  If the chosen trial wave function is close
to the exact ground-state wave function, the local energy is nearly
constant and the statistical efficiency of the algorithm is optimized.

If $f({\bf R},\tau)$ is constrained to be positive,
Eq.~(\ref{eq:fits}) may be interpreted as describing the time
evolution of the density of a population of ``random walkers''
multiplying or dying out as they diffuse and drift through a
$3N$-dimensional ``configuration space''.  The constraint that $f$ is
positive is known as the fixed-node approximation,~\cite{anderson}
because it forces the nodal surface of $\Psi({\bf R}, \tau)$ to be the
same as that of $\Phi_{\rm T}({\bf R})$.  In practice, the
distribution $f$ is represented by a few hundred walkers propagating
stochastically according to rules derived from Eq.~(\ref{eq:fits}).
The growth or decay rate of the total number of walkers in the
simulation depends on the average value of $E_{\rm L}({\bf R}) -
E_{\rm S}$, and the constant energy shift $E_{\rm S}$ is chosen to
ensure that the population remains stable on average.

The initial walker positions are normally picked from the probability
distribution $f({\bf R},\tau=0) = (\Phi_{\rm T}({\bf R}))^2$,
resulting in an initial population scattered throughout the
configuration space.  The nodal surface of $\Phi_{\rm T}({\bf R})$
divides this space into different nodal pockets, among which the
walkers are distributed.  During the simulation, the walkers never
cross the fixed nodal surface separating one pocket from another, and
so the fixed-node DMC algorithm proceeds independently in each pocket.

In the long-time limit, the probability density $f$ of the walkers
within nodal pocket $v_{\alpha}$ becomes proportional to
$\phi_{\alpha}({\bf R})\Phi_{\rm T}({\bf R})$, where the ``pocket
ground state'' $\phi_{\alpha}({\bf R})$ is the lowest energy real
normalized wave function which is zero outside $v_{\alpha}$ and
satisfies the fixed-node boundary conditions on the surface of
$v_{\alpha}$.  This function generally has gradient discontinuities
across the surface of $v_{\alpha}$, and the action of the kinetic
energy operator on these discontinuities produces delta function terms
which will be denoted $\delta_{\alpha}$.  The pocket ground state
therefore satisfies the equations:
 
\begin{eqnarray}
\label{eq:eigensoln}
\left. \begin{array}{c}
\hat{H}\phi_{\alpha}({\bf R}) = \epsilon_{\alpha}\phi_{\alpha}
({\bf R}) + \delta_{\alpha} \\
\phi_{\alpha}({\bf R})\Phi_{\rm T}({\bf R}) \geq 0
\end{array} \right\}
& \;\;\; \rm{when} \; {\bf R} \in \; {\it v}_{\alpha}
\nonumber \\
\phi_{\alpha}({\bf R}) = 0 \;\;\;\;\;\;\;
& \;\;\;\;\; \rm{when} \; {\bf R} \not\in \;
{\it v}_{\alpha} \;.
\end{eqnarray}
 
\noindent Since the walker density is positive, the sign of
$\phi_{\alpha}({\bf R})$ within $v_{\alpha}$ is the same as that of
$\Phi_{\rm T}({\bf R})$, implying that $\phi_{\alpha}({\bf R})$ and
$\Phi_{\rm T}({\bf R})$ have a non-zero overlap.  The form of
$\phi_{\alpha}({\bf R})$ does not depend on the details of $\Phi_{\rm
T}({\bf R})$, but only on the shape of its nodal surface.  The energy
given by the DMC simulation in nodal pocket $v_{\alpha}$ is equal to
the pocket eigenvalue $\epsilon_{\alpha}$.  This need not equal the
exact ground-state eigenvalue $E_0$ unless the trial nodal surface is
the same as that of the exact ground state $\Psi_0({\bf R})$, in which
case $\phi_{\alpha}({\bf R})$ is proportional to $\Psi_0({\bf R})$
within $v_{\alpha}$.

These results were used by Reynolds {\it et al.}~\cite{reynolds} and
Moskowitz {\it et al.}~\cite{moskowitz} to prove that the fixed-node
DMC energy is greater than or equal to the exact ground-state
energy. Reynolds' proof starts from the solution in a single nodal
pocket, and uses the permutations $P$ that do not flip spins to
construct a real wave function antisymmetric with respect to
interchanges of electron coordinates of the same spin,

\begin{equation}
\label{eq:antiwave}
\bar{\Psi}_{\alpha}({\bf R}) = \frac{1}{N_{\uparrow}!N_{\downarrow}!}
\sum_P (-1)^P \phi_{\alpha} (P{\bf R}) \equiv 
\hat{\mathcal{A}}\phi_{\alpha}({\bf R})\;.
\end{equation}

\noindent 
The function $\bar{\Psi}_{\alpha}({\bf R})$ cannot equal zero
everywhere since, at any point ${\bf R}$, all terms contributing to
the sum in Eq.~(\ref{eq:antiwave}) have the same sign.  This follows
from the antisymmetry of the trial function $\Phi_{\rm T}({\bf R})$,
which guarantees that all permutations $P$ for which the point $P{\bf
R}$ lies in nodal pocket $v_{\alpha}$ have the same parity (positive
if the sign of the trial function at ${\bf R}$ is the same as the sign
of the trial function in $v_{\alpha}$, negative otherwise).  The
function $\bar{\Psi}_{\alpha}({\bf R})$ is however equal to zero
everywhere on the trial nodal surface.  This can be deduced from the
knowledge that $\phi_{\alpha}({\bf R})$ is equal to zero there (as
well as in many other places), and that the nodal surfaces of
antisymmetric functions such as $\Phi_{\rm T}({\bf R})$ are not
altered by permutations.

The real antisymmetric function $\bar{\Psi}_{\alpha}({\bf R})$ is now
substituted into the standard quantum mechanical variational principle
to give,

\begin{eqnarray}
\label{eq:var} 
E_0 & \leq & \frac{\int \bar{\Psi}_{\alpha} {\hat{H}}
\bar{\Psi}_{\alpha}\, d{\bf R}}{\int \bar{\Psi}_{\alpha}
\bar{\Psi}_{\alpha}\, d{\bf R}} \, = \,
\frac{\int \bar{\Psi}_{\alpha} {\hat{H}}
{\hat{\mathcal A}} {\phi}_{\alpha}\, 
d{\bf R}}{\int \bar{\Psi}_{\alpha}
{\hat{\mathcal A}} 
{\phi}_{\alpha}\, d{\bf R}} \nonumber \\ 
& = &
\frac{\int \bar{\Psi}_{\alpha} {\hat{H}}
{\phi}_{\alpha}\, d{\bf R}}{\int \bar{\Psi}_{\alpha}
{\phi}_{\alpha}\, d{\bf R}} \, = \, \epsilon_{\alpha} \; ,
\end{eqnarray}

\noindent
where we have used the fact that $\hat{\mathcal A}$ commutes with
$\hat{H}$, that it is self-adjoint, and that it is idempotent (so that
$\hat{\mathcal A}\bar{\Psi}_{\alpha}({\bf R}) =
\bar{\Psi}_{\alpha}({\bf R})$).  The delta function terms appearing in
$\hat{H} \phi_{\alpha}$ do not contribute to the energy expectation
value because they occur on the fixed nodal surface where
$\bar{\Psi}_{\alpha}({\bf R})=0$.  If the nodal surface of $\Phi_{\rm
T}({\bf R})$ is the same as the nodal surface of the exact ground
state, the equality holds and the pocket eigenvalue
$\epsilon_{\alpha}$ is equal to $E_0$; but if $\Phi_{\rm T}({\bf R})$
does not have the correct nodal surface then $\epsilon_{\alpha} >
E_0$.  The energy $\epsilon_{\alpha}$ produced by the DMC simulation
in nodal pocket $v_{\alpha}$ is therefore minimized and equal to the
exact ground-state energy when the trial nodal surface is exact.
Since $\epsilon_{\alpha}$ is also expected to depend smoothly 
on the shape of the nodal surface, it follows that the error in 
$\epsilon_{\alpha}$ is in general second order in the error in the 
nodal surface.

For systems with reasonably well behaved local potentials and real
ground-state wave functions, the tiling theorem~\cite{tiling} states
that all the ground-state nodal pockets are related by permutation
symmetry.  A derivation of this theorem appears in
Sec.~\ref{sec:tiling}.  The tiling theorem holds even when the ground
state is degenerate, in which case every possible real linear
combination of the degenerate ground states possesses the tiling
property.  Many DMC simulations use trial wave functions with the same
nodal surface as the density-functional ground state.  Since the
density-functional Hamiltonian has a local potential, such trial
states always satisfy the tiling theorem.  This guarantees that the
value of $\epsilon_{\alpha}$ is the same in every nodal pocket, and
hence that the energy obtained in the fixed-node DMC simulation cannot
depend on how the walkers are distributed among the pockets.
 
In cases when the trial function does not possess the tiling property,
the walker population grows most rapidly in nodal pockets with low
values of $\epsilon_{\alpha}$ (or, equivalently, low average values of
$E_{\rm L}({\bf R})-E_{\rm S}$).  Since $E_{\rm S}$ is chosen such
that the total walker population stays roughly constant, and since
fixed-node DMC walkers never cross from one nodal pocket into another,
the walkers in less favorable pockets are gradually annihilated.
Although many different nodal pockets may have contained walkers
initially, the population becomes more and more concentrated in the
pocket or pockets with the lowest value of $\epsilon_{\alpha}$, and it
becomes more and more likely that only these pockets will be occupied.
The DMC energy is therefore almost certain to converge to the lowest
of the pocket eigenvalues of the nodal volumes initially occupied with
walkers, $E_{\rm DMC} = \rm{min}_{\alpha}(\epsilon_{\alpha})$.  This
means that the energy obtained in a DMC simulation which does not
possess the tiling property may depend on the initial distribution of
walkers.

\section{Absence of a General Variational Principle for Fixed-Node 
DMC Calculations of Excited States}
\label{sec:GenVar}

In a few cases, the exact nodal surface of an excited-state wave
function can be determined using symmetry arguments alone.  An example
is the first excited state of a particle in a one-dimensional square
well, which has a single nodal point located at the well center.  The
trial wave function can then be chosen to have exactly the same nodal
surface as the excited state, and the fixed-node DMC simulation gives
the exact excited-state eigenvalue.  This result holds whether or not
the trial wave function is orthogonal to all the lower energy
eigenstates.

In practice, however, the exact nodal surface is rarely determined by
symmetry alone, and it is rarely possible to choose a trial function
with exactly the same nodal surface as the excited state of interest.
Furthermore, the wave function of an arbitrary excited state need not
possess the tiling property, and so the DMC energy may depend on the
initial distribution of the walkers.
 
As an example, consider a hydrogen atom in its $2{\rm s}$ state,
$\Psi_{2{\rm s}}(r)$, with eigenvalue $E_{2{\rm s}}$.  The exact nodal
surface is a sphere of radius $r_0$, the value of which cannot be
determined using symmetry arguments.  If the imposed nodal surface is
exact, the pocket eigenvalues in the inner and outer pockets will both
be exactly equal to $E_{2{\rm s}}$; but if the nodal surface is a
sphere of radius $a \neq r_0$, the pocket eigenvalue $\epsilon_>$ of
the wave function $\Psi_>$ in the outer pocket will not equal the
pocket eigenvalue $\epsilon_<$ of the wave function $\Psi_<$ in the
inner pocket.

Consider the case when the fixed node is too close to the nucleus, $a
< r_0$.  According to the variational principle applied to the outer
pocket, the value of $\epsilon_>$ must be bounded above by the energy
expectation value of the trial function,

\begin{equation}
\Phi_{\rm T}^{>}(r) = \left \{ \begin{array}{ll} 0 \;\;\;\;\;\; & a
\leq r \leq r_0 \\ \Psi_{2{\rm s}}(r) & r > r_0 \; .  \end{array}
\right .
\end{equation}

\noindent 
This expectation value is the exact $2{\rm s}$ eigenvalue $E_{2{\rm
s}}$.  Clearly, we could construct a lower energy trial function by
removing the kink in $\Phi_{\rm T}^{>}(r)$ at $r = r_0$, and hence
$\epsilon_> < E_{2{\rm s}}$.  Similarly, we can take the pocket ground
state from the inner pocket, $\Psi_<$, and use it as a variational
trial function within the exact $2{\rm s}$ nodal surface $r = r_0$,

\begin{equation}
\Phi_{\rm T}^{<}(r) = \left \{ \begin{array}{ll} \Psi_<(r)
\;\;\; & r < a \\ 0 & a \leq r \leq r_0 \; .  \end{array} \right .
\end{equation}

\noindent 
In this case the energy expectation value of the trial function is
$\epsilon_<$, while the minimum possible expectation value for a wave
function with a node at $r_0$ is $E_{2{\rm s}}$.  Again, we could
construct a lower energy trial function by removing the kink at $r =
a$, and hence $\epsilon_< > E_{2{\rm s}}$.  Note that variational
arguments like these can be used whenever the exact nodal pocket
completely encloses the trial pocket or vice-versa, irrespective of
dimension or symmetry.  This will prove useful in
Sec.~\ref{sec:sepexample}.

The last paragraph showed that if $a \! < \! r_0$ then $\epsilon_> <
E_{2{\rm s}} < \epsilon_<$; if $a > r_0$, a similar derivation gives
$\epsilon_< < E_{2{\rm s}} < \epsilon_>$.  The two pockets have
different energies unless $a = r_0$.  How does this affect the
fixed-node DMC algorithm?  As long as the lower energy pocket contains
plenty of walkers initially, the walker population in the higher
energy pocket will almost certainly die out.  The fixed-node DMC
energy will then tend to the pocket eigenvalue of the lower energy
pocket, which is always less than or equal to $E_{2{\rm s}}$.  The DMC
energy is therefore \emph{maximized} when $a = r_0$.  If $a$ is
increased through $r_0$, the $\tau \rightarrow \infty$ DMC walker
population switches from the outer nodal pocket to the inner one, and
the slope of the graph of the DMC energy versus $a$ changes
discontinuously.  It follows that the error in the fixed-node DMC
energy is first order in the error in the nodal surface, not second
order as it is for the ground state.  This example shows that there is
no variational principle when fixed-node DMC is used to study general
excited states: the error in the DMC energy of the excited state may
increase linearly with the error in the nodal surface; and the DMC
energy need not be minimized when the nodal surface is exact.

\section{The Lowest Energy Eigenstate of Each Symmetry}
\label{sec:Lowest}

We now address the question of whether there is a variational
principle for fixed-node DMC simulations of the lowest energy state of
each symmetry.  We denote by ${\mathcal G}$ the group of spatial
transformations ${\mathcal T}$ (combinations of rotations,
reflections, translations, and inversions of all electrons
simultaneously) which leave the many-electron Hamiltonian
invariant. For simplicity, we consider this group to be
finite~\cite{lie} and of order $g$.  The full symmetry group of the
Hamiltonian is the direct product of ${\mathcal G}$ with the
permutation and time-reversal groups.  All the symmetry arguments in
this paper can easily be recast in terms of the full symmetry group
instead of ${\mathcal G}$, but real many-electron wave functions have
such simple time-reversal and permutation symmetries (they transform
according to the one-dimensional identity representation of the
time-reversal group and the one-dimensional antisymmetric
representation of the permutation group) that this is unnecessary.
Since the arguments based on the spatial group ${\mathcal G}$ are
somewhat easier to grasp, we choose to work with this group in what
follows.

We begin by showing that a symmetry-constrained variational principle
holds whenever the chosen trial function is real and has an invariant
nodal surface.  This theorem was stated without proof in a paper by
Caffarel and Claverie.~\cite{Caffarel} Here we sketch a proof.

A real trial function $\Phi_{\rm T}({\bf R})$ is said to have an
invariant nodal surface if the transformed function,

\begin{equation}
  \hat{Q}({\mathcal T})\Phi_{\rm T}({\bf R}) \equiv
  \Phi_{\rm T}({\mathcal T}^{-1}{\bf R}) \; ,
\end{equation}

\noindent
has the same nodal surface as $\Phi_{\rm T}({\bf R})$ for all
coordinate transformations ${\mathcal T} \in {\mathcal G}$.  Note that
the nodal surface is defined as the surface on which $\Phi_{\rm
T}({\bf R})$ is zero and across which it changes sign; $\Phi_{\rm
T}({\bf R})$ may also have additional zeros where the sign does not
change, but these are not on the nodal surface and need not be
invariant.

The proof relies on the properties of the function $\chi({\bf R})$
defined by,

\begin{equation}
\chi({\bf R}) = \left \{ \begin{array}{rl} 
0  & {\rm on\;the\;trial\;nodal\;surface} \\
+1 & {\rm in\;nodal\;pockets\;where\;}
\Phi_{\rm T} \geq 0 \\
-1 & {\rm in\;nodal\;pockets\;where\;}
\Phi_{\rm T} \leq 0 \; .
\end{array} \right .
\end{equation}  

\noindent 
Given any point ${\bf R}$ not on the nodal surface, and any spatial
transformation ${\mathcal T} \in {\mathcal G}$, it is clear that
$\chi({\bf R}) = \eta \chi({\mathcal T}^{-1}{\bf R})$, where $\eta=\pm
1$.  Furthermore, since the functions $\chi({\bf R})$ and
$\chi({\mathcal T}^{-1}{\bf R})$ have the same nodal surface and so
change sign together as ${\bf R}$ changes, the sign of $\eta$ must be
independent of ${\bf R}$.  This shows that all symmetries ${\mathcal
T} \in {\mathcal G}$ either leave $\chi({\bf R})$ unchanged or
multiply it by $-1$, from which it follows that $\chi({\bf R})$
transforms according to a one-dimensional irreducible representation
$\Gamma^{r}_{\mathcal G}$ of ${\mathcal G}$.

We can now adapt the proof of the ground-state variational principle
given in Sec.~\ref{sec:DMC}.  Take the pocket ground state
$\phi_{\alpha}$ and antisymmetrize it to obtain a function
$\bar{\Psi}_{\alpha}$ as in Eq.~(\ref{eq:antiwave}).  Now apply the
group theoretical projection operator,~\cite{Cornwell}

\begin{equation}
\label{eq:proj}
\hat{\mathcal P}^{r} = \frac{1}{g} \sum_{{\mathcal T} \in
\mathcal{G}} \Gamma^{r}_{\mathcal G}({\mathcal T}) 
\hat{Q}({\mathcal T}) \; ,
\end{equation}

\noindent
where $\Gamma^r_{\mathcal G}({\mathcal T})$ is the one-by-one matrix
representing ${\mathcal T}$.  The application of $\hat{\mathcal
P}^{r}$ produces a new antisymmetric state $\bar{\Psi}_{\alpha}^r$
which transforms according to the one-dimensional irreducible
representation $\Gamma^r_{\mathcal G}$.  Since the original pocket
ground state $\phi_{\alpha}({\bf R})$ had a non-zero overlap with
$\chi({\bf R})$, which is itself an antisymmetric function of symmetry
$\Gamma^r_{\mathcal G}$, the antisymmetric $\Gamma^r_{\mathcal G}$
component projected out by applying $\hat{\mathcal A}$ and then
$\hat{\mathcal P}^{r}$ to ${\phi}_{\alpha}$ cannot be zero.

The energy expectation value of $\bar{\Psi}_{\alpha}^r$ is greater
than or equal to the energy $E_0^r$ of the lowest antisymmetric state
of symmetry $\Gamma^r_{\mathcal G}$,

\begin{eqnarray}
\label{eq:Ealphap} 
E^{r}_0 & \leq & \frac{\int \bar{\Psi}_{\alpha}^{r*}{\hat{H}}
\bar{\Psi}_{\alpha}^{r}\, d{\bf R}}{\int \bar{\Psi}_{\alpha}^{r*}
\bar{\Psi}_{\alpha}^{r}\, d{\bf R}} \, = \,
\frac{\int \bar{\Psi}_{\alpha}^{r*}{\hat{H}}
\hat{\mathcal P}^{r} {\hat{\mathcal A}} {\phi}_{\alpha}\, 
d{\bf R}}{\int \bar{\Psi}_{\alpha}^{r*}
\hat{\mathcal P}^{r} {\hat{\mathcal A}} 
{\phi}_{\alpha}\, d{\bf R}} \nonumber \\ & = &
\frac{\int \bar{\Psi}_{\alpha}^{r*}{\hat{H}}
{\phi}_{\alpha}\, d{\bf R}}{\int \bar{\Psi}_{\alpha}^{r*}
{\phi}_{\alpha}\, d{\bf R}} \, = \, \epsilon_{\alpha} \; .
\end{eqnarray}

\noindent
In analogy with the ground-state proof of Sec.~\ref{sec:DMC}, these
manipulations rely on the self-adjointness and idempotency of the
operators $\hat{\mathcal A}$ and $\hat{\mathcal P}^r$, both of which
commute with $\hat{H}$.  It is also important that
$\bar{\Psi}_{\alpha}^{r*}({\bf R})$ is equal to zero everywhere on the
surface of $v_{\alpha}$, so that the delta function terms in the
expression for $\hat{H} {\phi}_{\alpha}({\bf R})$ from
Eq.~(\ref{eq:eigensoln}) do not contribute to the expectation value.
This is guaranteed because the fixed nodal surface is invariant.

Note that the symmetry of the trial function played no part in this
derivation; the only thing that mattered was the invariance of the
trial nodal surface.  In most DMC simulations, however, the trial
function does have a definite symmetry, and so transforms according to
a specific irreducible representation of ${\mathcal G}$.  Given that
functions transforming according to different irreducible
representations are orthogonal, and that the definition of $\chi({\bf
R})$ ensures that $\langle \chi | \Phi_{\rm T} \rangle > 0$, it
follows that any trial function with a definite symmetry and an
invariant nodal surface must transform according to the same
one-dimensional irreducible representation as the corresponding
$\chi({\bf R})$.  This implies that a real trial function transforming
according to an irreducible representation of dimension greater than
one cannot have an invariant nodal surface.

The converse is also true: any real trial function $\Phi_{\rm T}({\bf
R})$ transforming according to a one-dimensional irreducible
representation must have an invariant nodal surface.  This follows
from the transformation law,

\begin{equation}
\hat{Q}({\mathcal T})\Phi_{\rm T}({\bf R}) = 
\Gamma^r_{\mathcal G}({\mathcal T}) \Phi_{\rm T}({\bf R}) \; ,
\end{equation}

\noindent
and the observation that the real normalized function $\Phi_{\rm
T}({\bf R})$ remains real and normalized under all transformations in
${\mathcal G}$.  The one-by-one matrix $\Gamma^r_{\mathcal
G}({\mathcal T})$ is therefore equal to $\pm 1$, and the nodal surface
of $\hat{Q}({\mathcal T})\Phi_{\rm T}({\bf R})$ is the same as that of
$\Phi_{\rm T}({\bf R})$.

Putting everything together, we can now conclude that whenever the
real trial function transforms according to a one-dimensional
irreducible representation of ${\mathcal G}$, the nodal surface is
invariant, and the DMC energy is greater than or equal to the energy
of the lowest exact eigenfunction with that symmetry.  This is the
symmetry-constrained variational principle mentioned in the
introduction.  If the trial function transforms according to an
irreducible representation of dimension greater than one, the nodal
surface cannot be invariant, and the delta functions produced when
$\hat{H}$ is applied to $\phi_{\alpha}$ need not all occur where
$\bar{\Psi}_{\alpha}^{r*}=0$.  The above proof of the
symmetry-constrained variational principle therefore breaks down.  In
Sec.~\ref{sec:sepexample} we give an example of a system with a trial
function transforming according to an irreducible representation
$\Gamma^r_{\mathcal G}$ of dimension $d_r = 2$ for which $E_{\rm DMC}
< E^{r}_0$.

The restriction of the variational principle to trial functions
transforming according to one-dimensional irreducible representations
can be understood in a very simple way.  The problem being solved in a
fixed-node DMC simulation is not the imaginary-time Schr\"{o}dinger
equation of Eq.~(\ref{eq:its}), but the imaginary-time Schr\"{o}dinger
equation subject to the additional boundary conditions specified by
the trial nodal surface.  As a result, the relevant symmetry group is
not the group ${\mathcal G}$ of symmetries of $\hat{H}$, but the
subgroup ${\mathcal G}_{\rm FN}$ that leaves both the Hamiltonian and
the fixed nodal surface (boundary conditions) invariant.  The
fixed-node eigenfunctions need only conform to the symmetries in
${\mathcal G}_{\rm FN}$, and their transformation properties should be
analyzed in terms of the irreducible representations of ${\mathcal
G}_{\rm FN}$, not ${\mathcal G}$.  If the real trial function
transforms according to a one-dimensional irreducible representation
of ${\mathcal G}$, the fixed nodal surface is invariant and the groups
${\mathcal G}$ and ${\mathcal G}_{\rm FN}$ are the same.  The
symmetry-constrained variational principle discussed in this section
does not apply unless this is the case.

\section{Irreducible Representations of Dimension Greater than One}
\label{sec:d>1}

Can anything be said when the trial function transforms according to
an irreducible representation of dimension greater than one?  Suppose
that the real trial function $\Phi_{{\rm T},l}^{r}({\bf R})$
transforms as the $l$th row of an irreducible representation
$\Gamma^r_{\mathcal G}$ with dimension $d_r > 1$.  We already know
that the nodal surface of $\Phi_{{\rm T},l}^{r}({\bf R})$ cannot be
invariant with respect to all the operations in ${\mathcal G}$, but it
may be invariant under a subset of those operations. Any such subset
forms a proper subgroup, ${\mathcal G}_{\rm FN}$, which in general
depends on the row index $l$.

As in Sec.~\ref{sec:Lowest}, we define the function $\chi({\bf R})$
which is equal to $+1$ in all nodal pockets where $\Phi_{{\rm
T},l}^{r}({\bf R}) \geq 0$, equal to $-1$ in all nodal pockets where
$\Phi_{{\rm T},l}^{r}({\bf R}) \leq 0$, and equal to zero everywhere
on the nodal surface of $\Phi_{{\rm T},l}^{r}({\bf R})$.  By
construction, $\chi({\bf R})$ transforms according to a
one-dimensional irreducible representation of ${\mathcal G}_{\rm FN}$
(although not of ${\mathcal G}$).  Since all the pocket ground states
$\phi_{\alpha}$ have non-zero overlaps with $\chi({\bf R})$, they all
have non-zero components of the same subgroup symmetry as $\chi({\bf
R})$.  We can therefore re-use the symmetry-projection argument
leading to Eq.~(\ref{eq:Ealphap}) to show that the fixed-node DMC
energy is greater than or equal to the eigenvalue of the lowest energy
exact eigenfunction with the same subgroup symmetry as $\chi({\bf
R})$.  This provides a rigorous variational principle, but the
subgroup ${\mathcal G}_{\rm FN}$ that leaves the trial nodal surface
invariant is usually small and its irreducible representations provide
a correspondingly limited set of symmetry labels.  The bounds obtained
are therefore weak and this variational principle is of little use
unless optimized as explained below.

If the nodes of $\Phi_{{\rm T},l}^{r}({\bf R})$ were exact one could
form a trial function from any real linear combination of different
rows and always obtain the same DMC energy.  When the nodes are not
exact, however, the DMC method breaks this $d_r$-fold degeneracy, and
the variational lower bound on the DMC energy depends on the precise
linear combination of rows chosen in constructing the trial wave
function.  This freedom can be exploited to improve the weak
variational principle.

The $d_r$ functions $\Phi_{{\rm T},l}^{r}({\bf R})$
$(l=1,2,\ldots,d_r)$ are a basis for the irreducible representation
$\Gamma^r_{\mathcal G}$ of ${\mathcal G}$, and hence transform into
linear combinations of each other under all coordinate transformations
${\mathcal T} \in {\mathcal G}$.  This implies that they also
transform into linear combinations of each other under all coordinate
transformations ${\mathcal T} \in {\mathcal G}_s$, where ${\mathcal
G}_s$ is any proper subgroup of ${\mathcal G}$.  The subset of the
matrices $\Gamma^r_{\mathcal G}({\mathcal T})$ for which ${\mathcal T}
\in {\mathcal G}_s$ is therefore a representation of ${\mathcal G}_s$.
This representation is not in general irreducible.

To enumerate the different possibilities, begin by determining all the
proper subgroups ${\mathcal G}_s$ of ${\mathcal G}$.  For each such
subgroup, consider the representation of ${\mathcal G}_s$ consisting
of the matrices $\Gamma^r_{\mathcal G}({\mathcal T})$ with $T \in
{\mathcal G}_s$.  This representation may be decomposed into its
irreducible components with respect to ${\mathcal G}_s$,
\begin{equation}
\label{eq:Compat}
\Gamma^{r}_{\mathcal G} = n_{q_1} \Gamma^{q_1}_{{\mathcal G}_s}
\oplus n_{q_2} \Gamma^{q_2}_{{\mathcal G}_s} \oplus \cdots \oplus
n_{q_m} \Gamma^{q_m}_{{\mathcal G}_s} \; ,
\end{equation}

\noindent 
where the positive integer $n_{q_i}$ is the number of times the
irreducible representation $\Gamma^{q_i}_{{\mathcal G}_s}$ of
${\mathcal G}_s$ appears, so $d_r = \sum_i n_{q_i} d_{q_i}$.  The
irreducible representations of ${\mathcal G}_s$ appearing in
Eq.~(\ref{eq:Compat}) are said to be compatible~\cite{Cornwell} with
$\Gamma^r_{\mathcal G}$.  Note that since the group containing only
the identity element is always a subgroup of ${\mathcal G}$, it is
always possible to find at least one subgroup for which the reduction
of $\Gamma^r_{\mathcal G}$ contains a one-dimensional irreducible
representation $\Gamma^{q_i}_{{\mathcal G}_s}$.

A trial function transforming as the $l$th row of $\Gamma^r_{\mathcal
G}$ may contain components along all the compatible representations of
${\mathcal G}_s$, but it is always possible to construct linear
combinations,
\begin{equation}
  \label{eq:lincom}
  \Phi_{\rm T}^{q_i}({\bf R}) = \sum_{l=1}^{d_r} c_l^{q_i} 
  \Phi_{{\rm T},l}^{r}({\bf R}) \; ,
\end{equation}
transforming according to each particular compatible representation
$\Gamma^{q_i}_{{\mathcal G}_s}$.  Every real function $\Phi_{\rm
T}^{q_i}({\bf R})$ corresponding to a one-dimensional representation
of ${\mathcal G}_s$ has an invariant nodal surface with respect to
${\mathcal G}_s$.  A DMC energy calculated using the nodes of such a
function therefore satisfies the variational principle $E_{\rm DMC}
\geq E^{q_i}_0$.

Clearly, the strength of the variational principle obtained depends on
the choices of the subgroup ${\mathcal G}_s$ and the one-dimensional
representation $\Gamma^{q_i}_{{\mathcal G}_s}$.  In some cases, one
can find a subgroup ${\mathcal G}_s$ with a one-dimensional
representation $\Gamma^{q_i}_{{\mathcal G}_s}$ which is compatible
with $\Gamma^r_{\mathcal G}$ but incompatible with all those
irreducible representations $\Gamma^{r'}_{\mathcal G}$ for which
$E_0^{r'}<E_0^r$.  A DMC energy calculated using the trial function
$\Phi_{\rm T}^{q_i}({\bf R})$ is then guaranteed to be greater than or
equal to $E_0^r$.  In general, if one knows the ordering of the energy
levels beforehand, one can use the compatibility analysis to find the
one-dimensional irreducible representation $\Gamma^{q_i}_{{\mathcal
G}_s}$ which gives the most stringent energy bound for the eigenvalue
of interest.

As an application of this symmetry-constrained variational principle,
consider a crystal with space group ${\mathcal G}$.  We wish to
establish whether it is possible to use fixed-node DMC to obtain a
variational estimate of the eigenvalue $E_{0}({\bf k})$ of the lowest
energy eigenstate with crystal momentum ${\bf k}$.  The relevant
subgroup ${\mathcal G}_s$ is the translation group, which is Abelian
and has only one-dimensional irreducible representations.  These are
labeled by the crystal momentum ${\bf k}$, and so the statement of the
symmetry-constrained variational principle is very straightforward: if
the DMC trial function has crystal momentum ${\bf k}$, the DMC energy
is greater than or equal to $E_0({\bf k})$.  It is worth noting that
this statement holds whether or not the lowest energy state of crystal
momentum ${\bf k}$ is degenerate.  In the degenerate case the trial
function transforms according to a multi-dimensional irreducible
representation of ${\mathcal G}$, but it still has crystal momentum
${\bf k}$ and so still transforms according to a one-dimensional
irreducible representation of the translation subgroup ${\mathcal
G}_s$.  This is sufficient to guarantee that the DMC energy is greater
than or equal to $E_0({\bf k})$.

The symmetry-constrained DMC variational principle for the lowest
energy state of crystal momentum ${\bf k}$ is unfortunately much less
useful than it appears, because most Bloch states are complex and 
cannot be used as fixed-node DMC trial functions.  Instead, the
standard approach is to use a real linear combination of a Bloch
function and its complex conjugate.  (This is justified by the
assumption of time-reversal invariance, which guarantees that if
$\Psi_{\bf k}$ is an eigenfunction with eigenvalue $E_0({\bf k})$ then
so is $\Psi^{*}_{\bf k}$.)  Since such trial functions contain
components with two different wave vectors, ${\bf k}$ and $-{\bf k}$,
they do not transform according to a single irreducible representation
of the translation group.  The symmetry-constrained variational 
theorem is therefore inapplicable, and it is possible that the DMC 
energy may lie below $E_0({\bf k})$.  In Sec.~\ref{sec:sepexample} we 
show by means of a specific example that such calculations may indeed 
produce a DMC energy which is lower than $E_0({\bf k})$.

An important exception arises when the wave vector ${\bf k}$ is equal
to half a reciprocal lattice vector, in which case ${\bf k}$ and
$-{\bf k}$ are alternative labels for the same irreducible
representation of the translation group.  The linear combination of
$\Psi_{\bf k}$ and $\Psi_{\bf k}^{*}$ is then a pure Bloch function,
and the normal proof of the symmetry-constrained variational principle
applies.  As long as ${\bf k}$ equals half a reciprocal lattice
vector, the DMC energy is greater than or equal to the energy of the
lowest exact eigenstate with crystal momentum ${\bf k}$.

\section{Generalizations of the Tiling Theorem}
\label{sec:tiling}

In his paper on fermion nodes,~\cite{tiling} Ceperley stated that the
tiling theorem could be generalized to the case where there were other
discrete symmetries present.  A more precise statement is that, given
a Hamiltonian with a reasonable local potential and a symmetry group
${\mathcal G}$, the tiling theorem applies to any real state which is
the lowest energy eigenfunction of a symmetry $\Gamma^r_{\mathcal G}$
with dimension $d_r = 1$.  This can be demonstrated via a simple
generalization of Ceperley's proof~\cite{tiling} of the ground-state
tiling theorem.

Consider the real antisymmetric state $\Psi_0^r({\bf R})$ which is the
lowest energy eigenfunction transforming according to the
one-dimensional irreducible representation $\Gamma_{\mathcal G}^r$.
This function may have many nodal pockets, but pick one at random and
color it blue.  Now apply a symmetry operator from the group
containing the spatial symmetries (rotations, translations,
reflections and inversions) and the permutations.  Since the nodal
surface is invariant, this symmetry operator maps the region of space
within the blue nodal pocket into itself or into one of the other
nodal pockets.  If the blue pocket is mapped into some other nodal
pocket, this pocket is equivalent to the blue one by symmetry and is
also colored blue.  Repeat this process for every operator in the
group, until all the pockets equivalent to the original one have been
found and colored blue.

There are now two possibilities: the blue regions may fill the entire
configuration space, in which case the nodal pockets are all
equivalent by symmetry and the tiling theorem holds; or there may be
other inequivalent nodal pockets which have not yet been found.  We
can rule out the second possibility using the following argument.

Assume that the blue nodal pockets do not fill the configuration
space, and that the local potential $V({\bf R})$ and all its
derivatives are finite except at the Coulomb singularities occurring
when two charged particles approach each other.  As long as the system
is not one-dimensional (in which case different arguments are
required), this ensures that almost every point on the nodal surface
lies a finite distance away from the nearest singularity in the
potential.  The eigenfunction $\Psi_0^r({\bf R})$ may therefore be
expanded as a power series with a finite radius of convergence about
almost any point ${\bf R}_s$ on the nodal surface.  If the gradient of
$\Psi_0^r({\bf R})$ is assumed to be zero over any finite area of the
nodal surface surrounding ${\bf R}_s$, it can be shown that every term
in this series has to be zero, and hence that $\Psi_0^r({\bf R}) = 0$
everywhere within the radius of convergence.  Since any solution of
the Schr\"{o}dinger equation can be analytically continued around the
isolated singularities in the potential, this further implies that
$\Psi_0^r({\bf R})$ is zero everywhere.  We therefore conclude that
the gradient of $\Psi_0^r({\bf R})$ must be non-zero almost everywhere
on the nodal surface.

Now consider the trial function $\tilde{\Psi}^r({\bf R})$, which is
defined to equal $\Psi_0^r({\bf R})$ within the blue pockets and zero
elsewhere.  This trial function is antisymmetric and transforms
according to the irreducible representation $\Gamma_{\mathcal G}^r$,
but has gradient discontinuities on the nodal surfaces separating the
blue pockets from the rest of configuration space. It satisfies the
Schr\"{o}dinger-like equation,

\begin{equation}
\label{eq:newdelta}
\hat{H}\tilde{\Psi}^r({\bf R}) = E_0^r \tilde{\Psi}^r({\bf R})
+ \delta^r  \;,
\end{equation}

\noindent
where the symbol $\delta^r$ denotes the delta functions produced by
the action of the kinetic energy operator on the gradient
discontinuities.  The delta functions occur where $\tilde{\Psi}^r = 0$
and so do not affect the energy expectation value,

\begin{equation}
E_0^r =
\frac{ \langle \Psi^r_0 | \hat{H} | \Psi^r_0 \rangle }
     { \langle \Psi^r_0 | \Psi^r_0 \rangle } =
\frac{ \langle \tilde{\Psi}^r | \hat{H} | \tilde{\Psi}^r \rangle }
     { \langle \tilde{\Psi}^r | \tilde{\Psi}^r \rangle } \; .
\label{eq:tilvar}
\end{equation}

\noindent
We know, however, that a state which has gradient discontinuities
almost everywhere on a finite area of the nodal surface cannot be an
eigenfunction unless the potential is infinite almost everywhere on
that area.  Since we are assuming that this is not the case, the
function $\tilde{\Psi}^r$ must contain excited-state components of
symmetry $\Gamma^r_{\mathcal G}$ and cannot have the same energy
expectation value as the lowest energy state of that symmetry.  This
conclusion contradicts Eq.~(\ref{eq:tilvar}), and so the assumption
that the blue nodal pockets do not fill the configuration space must
have been incorrect.  All the nodal pockets of $\Psi_0^r$ are
therefore equivalent by symmetry.

This proves the tiling theorem for any real state which is the lowest
energy eigenfunction of a symmetry $\Gamma_{\mathcal G}^r$ with
dimension $d_r = 1$.  An obvious corollary is that there is also a
tiling theorem for the lowest energy state transforming according to
any one-dimensional irreducible representation of any subgroup
${\mathcal G}_s$ of ${\mathcal G}$.  This statement is analogous to
the weaker variational principle discussed in Sec.~\ref{sec:d>1}.  In
every case when we have demonstrated the existence of a DMC
variational principle, we have therefore also been able to demonstrate
the existence of a tiling theorem.  Our analysis has been restricted
to the physically interesting case of a local potential which is
finite everywhere except at Coulomb singularities, but our conclusions
may be somewhat more general than this suggests.

The familiar many-fermion ground-state tiling theorem may be viewed as
a special case of the subgroup tiling theorem mentioned above.  The
permutation group is always a subgroup of the full symmetry group
(which contains both spatial and permutation symmetries), and the
many-fermion ground state is the lowest energy state which transforms
according to the one-dimensional antisymmetric irreducible
representation of that subgroup.  The subgroup tiling theorem
therefore guarantees that the many-fermion ground state possesses the
tiling property.  Note that the tiling property holds with respect to
the permutation subgroup, not the full symmetry group.  This means
that it is only the elements of the permutation subgroup that need be
applied to the initial blue pocket to find all equivalent pockets and
turn the whole configuration space blue.

The above derivation of the subgroup tiling theorem only applies to
states that transform according to one-dimensional irreducible
representations of the chosen subgroup ${\mathcal G}_s$.  Such states
may also transform according to multi-dimensional irreducible
representations of the full symmetry group ${\mathcal G}$, so the
tiling theorem is not restricted to non-degenerate energy levels.  In
systems with degenerate many-fermion ground states, for example, any
real linear combination of the degenerate ground states is
antisymmetric and so possesses the tiling property with respect to the
permutation subgroup.  This result holds even though the nodal surface
is not invariant under all the operators from the full symmetry group.

The most important consequence of the generalized tiling theorem is
exactly as in the ground-state case.  It is common for a fixed-node
DMC trial wave function to have the same nodal surface as an energy
eigenfunction calculated using an approximate method such as
local-density-functional theory.  The approximate Hamiltonian is
chosen to have the same symmetries as the exact Hamiltonian, but may
also have extra symmetries which are not relevant to the argument and
may be ignored.  If the approximate Hamiltonian has a reasonably well
behaved local potential, its eigenstates have the same tiling
properties with respect to ${\mathcal G}$ as the corresponding exact
eigenstates.  The lowest energy eigenstate transforming according to
any one-dimensional irreducible representation of ${\mathcal G}$ or
any subgroup of ${\mathcal G}$ therefore satisfies the tiling theorem.
This ensures that the energy produced by a DMC simulation using the
nodal surface of such a state is independent of the initial walker
distribution.

\section{Tight-Binding Example}
\label{sec:tbexample}

This section and the next describe simple examples that show what
happens when a trial wave function transforming according to an
irreducible representation with $d_r\!>\!1$ is used to define the
trial nodal surface for a DMC simulation.  We find that the DMC energy
may indeed be lower than the eigenvalue of the lowest exact eigenstate
with the same symmetry as the trial function.  This demonstrates that
the variational principle of Eq.~(\ref{eq:Ealphap}) does not apply
when $d_r\!>\!1$.

In searching for a suitable example system, we found it convenient to
impose the following restrictions: (i) the group of the Hamiltonian is
finite; (ii) the trial wave function and its nodal surface can be
easily visualized; and (iii) the exact eigenstates of the Hamiltonian
and the pocket ground states of Eq.~(\ref{eq:eigensoln}) can be
calculated without numerical error.  These restrictions are
unnecessary, but make the analysis much simpler.  To satisfy criterion
(ii), we must choose a system containing at most three electrons in
one dimension or a single electron in two or three dimensions.  The
nodal surfaces of one-dimensional systems are not easily altered
because they are almost entirely determined by the antisymmetry of the
wave function, so we decided to concentrate on one-electron systems in
two and three spatial dimensions.

The first system we studied was an electron confined to a rectangular
box in two or three dimensions.  Although this system does have
multi-dimensional irreducible representations, we failed to find an
example in which the DMC energy calculated using a trial state of a
given symmetry was less than the eigenvalue of the lowest energy
eigenstate of that symmetry.  The second system we tried was more
successful.  We start by discussing a simple tight-binding realization
of this system, and then in Sec.~\ref{sec:sepexample} present an
alternative realization based on a separable solution of the
Schr\"{o}dinger equation.

Consider a molecule containing one electron moving in the potential of
three protons fixed at the corners of an equilateral triangle.  The
symmetry group of the Hamiltonian of this system is called $D_{3{\rm
h}}$.~\cite{Cornwell} A convenient way to generate DMC trial wave
functions with specific symmetries is to solve the Schr\"{o}dinger
equation within a tight-binding approximation using a single
spherically-symmetric atomic-like orbital $\xi(|{\bf r}|)$ centered on
each proton.  Once a tight-binding eigenfunction has been found and
used to define a trial nodal surface, the fixed-node DMC algorithm can
be used to solve the original Hamiltonian exactly subject to the
fixed-node constraint.

The ground state of the molecule is a non-degenerate nodeless function
transforming according to the identity representation.  Some of the
excited-state eigenfunctions must be doubly degenerate, however, since
the symmetry group $D_{3{\rm h}}$ has two two-dimensional irreducible
representations, one of which is called $\Gamma^3_{D_{3{\rm h}}}$.

The tight-binding Hamiltonian has only three eigenstates, the nodeless
ground state and a degenerate pair of excited states which can be
written as the Bloch functions,

\begin{eqnarray}
\Psi_{+}({\bf r}) & = & \sum_{j=0}^{2} \xi(|{\bf r} - {\bf d}_j |)
e^{i2\pi j/3} \;, \nonumber \\ 
\Psi_{-}({\bf r}) & = & \sum_{j=0}^{2}
\xi(|{\bf r} - {\bf d}_j |) e^{-i2\pi j/3}\;,
\end{eqnarray}

\noindent where ${\bf d}_0$, ${\bf d}_1$, and ${\bf d}_2$ are the
position vectors of the three protons.  These two Bloch functions
transform into linear combinations of each other under the operations
of the point group (there is no need to include time-reversal
symmetry) and form a basis for $\Gamma^3_{ D_{3{\rm h}}}$. Taking
linear combinations of $\Psi_{+}$ and $\Psi_{-}$, one can form the
real functions,

\begin{eqnarray}
\label{eq:psi1_psi2}
\Psi_{1}({\bf r}) & = & \xi(|{\bf r} - {\bf d}_1|) - 
\xi(|{\bf r} - {\bf d}_2|) \; , \nonumber \\
\Psi_{2}({\bf r}) & = & 2\xi(|{\bf r} - {\bf d}_0|) 
- \xi(|{\bf r} - {\bf d}_1|) - \xi(|{\bf r} - {\bf d}_2|) \; ,
\end{eqnarray}

\noindent
which form an alternative basis for the same irreducible
representation.

\begin{figure}
\begin{center}
\leavevmode
\epsfxsize=8cm \epsfbox{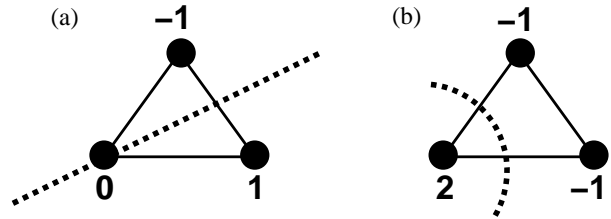}
\end{center}
\caption{ The single-electron trial states (a) $\Psi_1$ and (b)
$\Psi_2$ from Eq.~(\ref{eq:psi1_psi2}).  The dashed lines show where
the nodal surfaces cross the plane of the molecule.  The expansion
coefficients of $\Psi_1$ and $\Psi_2$ in terms of the tight-binding
basis functions are also shown. }
\label{fig1}
\end{figure}

Fig.~(\ref{fig1}) shows the nodal surfaces of $\Psi_1$ and $\Psi_2$
along with the expansion coefficients from Eq.~(\ref{eq:psi1_psi2}).
The nodal surface of $\Psi_1$ does not depend on the particular
spherically-symmetric tight-binding basis functions chosen and turns
out to be an exact excited-state nodal surface.  The nodal surface of
$\Psi_2$ is not exact, however, and its precise shape depends on the
details of the atomic-like orbitals used in the tight-binding
model. Different choices of $\xi(|{\bf r}|)$ give trial functions with
the same symmetries but different nodal surfaces.  By changing
$\xi(|{\bf r}|)$, it is possible to change the relative sizes of the
two nodal pockets of $\Psi_2$, causing one to grow at the expense of
the other.  As in the example of the $2{\rm s}$ state of the H atom
discussed earlier, this suggests that in many cases the pocket
eigenvalue of one of the two nodal pockets will be too high while that
of the other is too low.  The DMC energy will then lie below the
energy of the exact eigenstate of interest.  In
Sec.~\ref{sec:sepexample} we numerically solve a specific example with
the same symmetry properties as this system and observe exactly this
behavior.

Here we apply the weaker variational principle described in the
previous section to the degenerate excited state with symmetry
$\Gamma^3_{ D_{3{\rm h}}}$ of the group $D_{3{\rm h}}$.  The largest
subgroups $C_{3{\rm h}}$ and $D_3$ each have six elements. If the
subgroup $D_3$ is used, the compatibility relation of
Eq.~(\ref{eq:Compat}) becomes,

\begin{equation}
\Gamma^3_{D_{3{\rm h}}} = \Gamma^3_{D_3} \;,
\end{equation}

\noindent
where $\Gamma^3_{D_3}$ is two dimensional. This result is not useful
since we are seeking a reduction that contains one-dimensional
representations.  Using the subgroup $C_{3{\rm h}}$ results in two
one-dimensional representations
 
\begin{equation}
\Gamma^3_{D_{3{\rm h}}} = \Gamma^5_{C_{3{\rm h}}} \oplus
\Gamma^6_{C_{3{\rm h}}} \;.
\end{equation}

\noindent
However, any trial wave function that transforms as either
$\Gamma^5_{C_{3{\rm h}}}$ or $\Gamma^6_{C_{3{\rm h}}}$ is complex and
therefore unsuitable for fixed-node DMC.  

If we use the smaller subgroup, $C_{2{\rm v}}$, which has four
elements, we obtain the compatibility relation

\begin{equation}
\Gamma^3_{D_{3{\rm h}}} = \Gamma^1_{C_{2{\rm v}}} \oplus
\Gamma^4_{C_{2{\rm v}}} \;,
\end{equation}

\noindent
from which we can construct real trial wave functions and apply the
weaker variational principle.  In fact, the trial functions $\Psi_1$
and $\Psi_2$ already have the correct transformation properties:
$\Psi_1$ transforms as the one-dimensional representation
$\Gamma^4_{C_{2{\rm v}}}$, while $\Psi_2$ transforms as the
one-dimensional representation $\Gamma^1_{C_{2{\rm v}}}$. The group
$C_{2{\rm v}}$ therefore preserves the nodal surfaces of $\Psi_1$ and
$\Psi_2$.  Because the symmetry corresponding to the representation
$\Gamma^1_{C_{2{\rm v}}}$ is compatible with the ground-state symmetry
as well as with $\Gamma^3_{D_{3{\rm h}}}$, a DMC simulation using the
trial function $\Psi_2$ satisfies only the weaker variational
principle $E_{\rm DMC} \geq E_0$, where $E_0$ is the overall
ground-state energy.  The representation $\Gamma^4_{C_{2{\rm v}}}$ is
not compatible with the ground-state symmetry, however, and so a
simulation using the trial function $\Psi_1$ gives a stronger
variational principle.  For the example studied in the next section,
it turns out that the $\Gamma^3_{D_{3{\rm h}}}$ state of interest is
the lowest energy exact eigenstate with which the representation
$\Gamma^4_{C_{2{\rm v}}}$ is compatible.  The strong variational
principle therefore applies and $E_{\rm DMC}\geq E_0^3$, where $E_0^3$
is the exact $\Gamma^3_{D_{3{\rm h}}}$ eigenvalue.  It is important to
appreciate that this is not a general result; it is not always
possible to pick a trial function that maintains the strong
variational principle for a given symmetry.

Note that the strong variational principle applies because the trial
wave function $\Psi_1$ transforms according to the irreducible
representation $\Gamma^4_{C_{2{\rm v}}}$ of the subgroup $C_{2{\rm
v}}$.  If we had chosen a different pair of trial functions,
constructed by taking linear combinations of $\Psi_1$ and $\Psi_2$,
both would have contained components along $\Gamma^4_{C_{2{\rm v}}}$
and $\Gamma^1_{C_{2{\rm v}}}$.  The only subgroup preserving the nodes
would then have been the group of the identity, and the only
variational principle would have been with respect to the overall
ground-state energy, $E_{\rm DMC} \geq E_0$.  This illustrates the
general rule that the strongest variational principles are obtained by
choosing trial functions which transform according to specific
one-dimensional irreducible representations of specific subgroups of
${\mathcal G}$.

\section{Separable Example}
\label{sec:sepexample}

We now present an explicit solution of a different example with the
same $D_{3{\rm h}}$ symmetry group as the triangular molecule
discussed above.  Consider a particle of unit mass moving in a
triangular potential in three dimensions.  The wave function
$\Psi({\bf r})$ obeys the Schr\"{o}dinger equation,

\begin{equation}
\label{eq:Hamil}
\left ( - \frac{1}{2} \nabla^2 + V({\bf r}) \right ) \Psi({\bf r}) = E
\Psi({\bf r}) \; ,
\label{eq:sepham}
\end{equation}

\noindent
where

\begin{equation}
\label{eq:Poten}
V({\bf r}) = \frac{\cos(3\theta)}{\rho^2} \;,
\end{equation}

\noindent
with $\rho$, $\theta$ and $z$ the usual cylindrical coordinates.  The
boundary conditions are $\Psi(\rho,\theta,z) = 0$ for $\rho \geq 1$ or
$|z| \geq \pi/2$, confining the particle within a cylinder.

Writing $\Psi({\bf r}) = R ( \rho ) \Theta ( \theta ) Z(z)$, the
Schr\"{o}dinger equation separates into:

\begin{eqnarray}
- \frac{1}{2} \frac{d^2 Z}{dz^2} & = & E_z Z \; ,
\label{eq:zeq}  \\
- \frac{1}{2} \frac{d^2 \Theta}{d \theta^2} + \cos ( 3\theta ) \Theta
& = & E_{\theta} \Theta \; ,
\label{eq:thetaeq}  \\
- \frac{1}{2\rho} \frac{d}{d \rho} \left ( \rho \frac{dR}{d\rho}
\right ) + \left ( E_z + \frac{E_{\theta}}{\rho^2} \right ) R & = & E
R \;\; .
\label{eq:rhoeq}
\end{eqnarray}

\noindent
The lowest energy eigenfunction of Eq.~(\ref{eq:zeq}) is

\begin{equation}
Z(z) = \sqrt{\frac{2}{\pi}} \cos(z) \;,
\end{equation}

\noindent
with eigenvalue $E_z = 1/2$. Eq.~(\ref{eq:rhoeq}) may be simplified by
the substitutions $r = \sqrt{2(E - E_z)} \rho$ and $\nu = \sqrt{2
E_{\theta}}$, which yield Bessel's equation,

\begin{equation}
r^2 \frac{dR}{dr^2} + r \frac{dR}{dr} + ( r^2 - \nu^2 ) R = 0 \; .
\end{equation}

\noindent
The general solutions are the Bessel and Neumann functions,
$J_{\nu}(r)$ and $N_{\nu}(r)$, but only $J_{\nu}(r)$ is well behaved
at the origin.  Hence

\begin{equation}
R(\rho) = J_{\sqrt{2E_{\theta}}} \left ( \sqrt{2(E - E_z)} \rho \right
) \; ,
\end{equation}

\noindent
with the energy $E$ determined by the boundary condition at $\rho =
1$,

\begin{equation}
J_{\sqrt{2E_{\theta}}} ( \sqrt{2(E - E_z )} ) = 0 \; .
\end{equation}

\noindent
Eq.~(\ref{eq:thetaeq}) can be transformed into Mathieu's
equation~\cite{Abram} by a simple change of variables, but here we
solve it numerically by expanding the eigenfunctions in (normalized)
sines and cosines,

\begin{equation}
  \Theta(\theta) = a_0 \frac{1}{\sqrt{2\pi}} + \sum_{n=1}^{M}
         \left ( a_n \frac{\cos(n \theta)}{\sqrt{\pi}}  + 
                 b_n \frac{\sin(n \theta)}{\sqrt{\pi}} \right ) \; ,
\end{equation}

\noindent
and diagonalizing the corresponding Hamiltonian matrix.  The results
converge rapidly with $M$, and choosing $M\!=\!50$ gives very accurate
eigenvalues for the lowest few eigenstates.

\begin{figure}
\begin{center}
\leavevmode
\epsfxsize=8cm \epsfbox{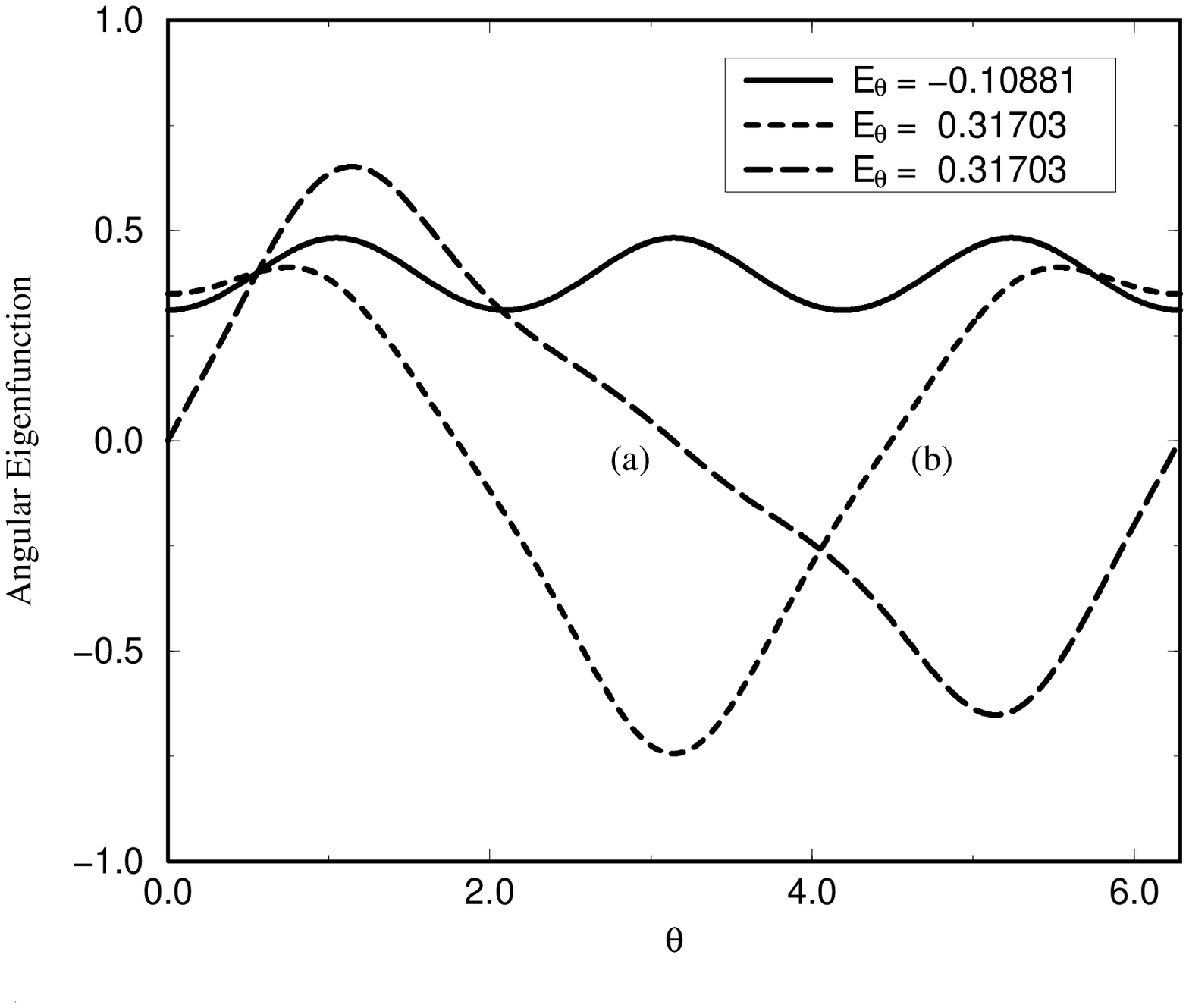}
\end{center}
\caption{ Angular eigenfunctions $\Theta(\theta)$ of the separable
Hamiltonian of Eq.~(\ref{eq:sepham}) with $V({\bf r}) = 
\cos(3\theta)/\rho^2$.  The functions (a) and (b) are analogous to 
the tight-binding states (a) and (b) shown in Fig.~(\ref{fig1}).  The
corresponding angular eigenvalues $E_{\theta}$ are also shown.}
\label{fig2}
\end{figure}

The three lowest energy angular eigenfunctions are shown in
Fig.~(\ref{fig2}).  As stated in Sec.~\ref{sec:tbexample}, the lowest
energy eigenstate is a nodeless function invariant under all elements
of the symmetry group of the Hamiltonian.  The next two states are a
degenerate pair forming a basis for the irreducible representation
$\Gamma^3_{{\rm D}_{3{\rm h}}}$.  One of the two has nodes at $\theta
= 0$ and $\theta = \pi$, and will be called $\Psi_1$ in analogy with
the corresponding state from Sec.~\ref{sec:tbexample}; the other has
nodes at $\theta = \pm 1.7934$ radians, and will be called $\Psi_2$.
Note that $\Psi_1$ is the lowest energy eigenfunction transforming
according to the one-dimensional irreducible representation
$\Gamma^4_{C_{2{\rm v}}}$ of the subgroup $C_{2{\rm v}}$; it therefore
possesses the tiling property with respect to $C_{2{\rm v}}$.  The
$\Gamma^1_{C_{2{\rm v}}}$ symmetry of $\Psi_2$ is shared by the
overall ground state, however, so $\Psi_2$ is not the lowest
eigenstate of that subgroup symmetry and does not satisfy a tiling
theorem.

\begin{figure}
\begin{center}
\leavevmode
\epsfxsize=8cm \epsfbox{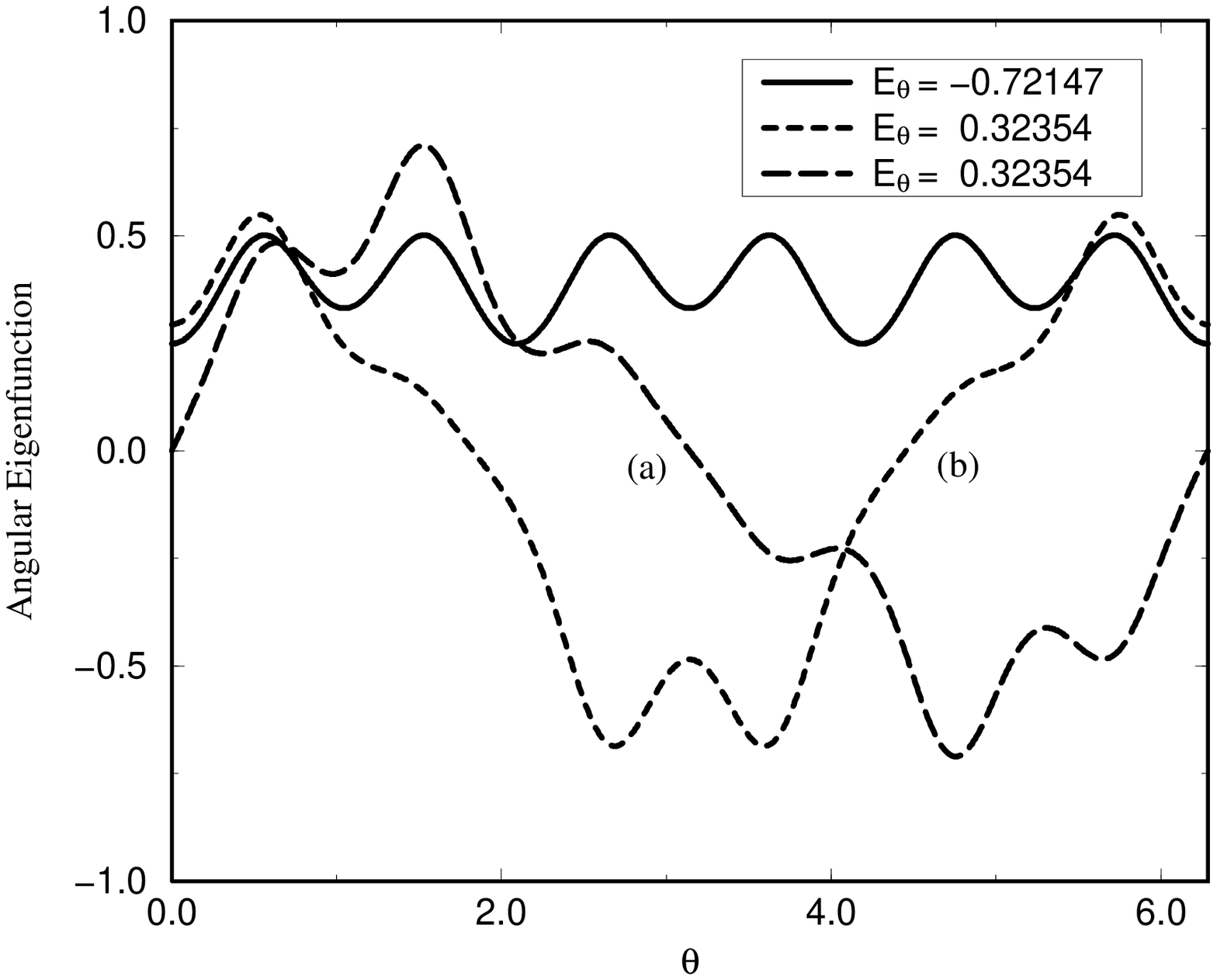}
\end{center}
\caption{ Angular eigenfunctions $\Theta(\theta)$ of the separable
Hamiltonian of Eq.~(\ref{eq:sepham}) with $V({\bf r}) = (\cos(3\theta)
+ 5 \cos(6\theta))/\rho^2$.  The functions (a) and (b) are analogous
to the tight-binding states (a) and (b) shown in Fig.~(\ref{fig1}).
The corresponding angular eigenvalues $E_{\theta}$ are also shown.}
\label{fig3}
\end{figure}

Consider how DMC might be used to find the eigenvalue of the lowest
energy $\Gamma^3_{{\rm D}_{3{\rm h}}}$ doublet.  We do not want to
impose the exact nodal surface since DMC would then give the exact
answer and we would learn little about variational principles, so we
seek a trial function with a different nodal surface but the same
symmetry.  We choose to generate such a trial function by solving the
Schr\"{o}dinger equation for a different triangular potential,

\begin{equation}
\hat{H} = - \frac{1}{2} \nabla^2 + \frac{\cos(3\theta)}{\rho^2} + 
          \mu \frac{\cos(6\theta)}{\rho^2} \; ,
\end{equation}

\noindent
where $\mu$ is an adjustable parameter.  The first three angular
eigenfunctions of this Hamiltonian when $\mu=5$ are shown in
Fig.~(\ref{fig3}).  Since the $\cos(6\theta)$ term does not change the
symmetry group, these eigenfunctions still belong to the same
irreducible representations.  However, the nodal angle of $\Psi_2$ has
moved slightly in response to the perturbation.  One of the two nodal
volumes of this trial wave function is therefore slightly too small,
while the other is slightly too large.

The trial nodal pocket that is too large (small) encloses (is enclosed
by) the corresponding exact nodal pocket.  We can therefore apply the
variational argument used in the discussion of the $2{\rm s}$ state of
the hydrogen atom to show that the pocket eigenvalue from the pocket
that is too small must be greater than the exact eigenvalue, while the
pocket eigenvalue from the pocket that is too large must be less than
the exact eigenvalue.  As in the H $2{\rm s}$ example, therefore, the
DMC energy is always $\leq$ the exact eigenvalue.  The maximum of the
DMC energy, equal to the exact $\Gamma^3_{{\rm D}_{3{\rm h}}}$
ground-state eigenvalue, is attained only when the nodal angle is
exact; and the slope of the graph of DMC energy against nodal angle
changes discontinuously at this point.  The error in the DMC energy is
first order, not second order, in the error in the nodal angle.

\begin{figure}
\begin{center}
\leavevmode
\epsfxsize=8cm \epsfbox{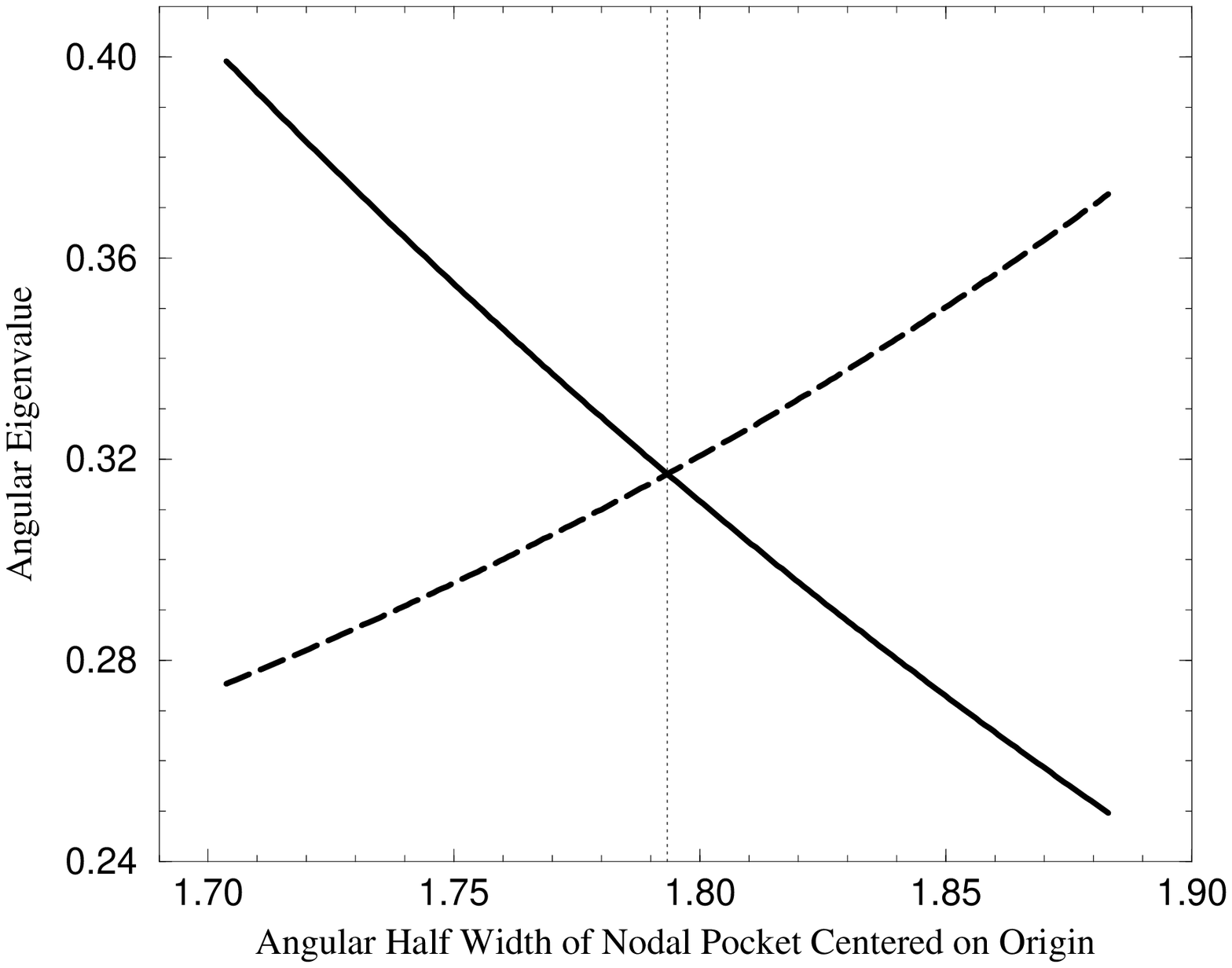}
\end{center}
\caption{ The angular pocket eigenvalues $E_{\theta}$ of the two nodal
pockets of the separable example from Sec.~\ref{sec:sepexample}.  The
Hamiltonian is that of Eq.~(\ref{eq:sepham}) with $V({\bf
r})=\cos(3\theta)/\rho^2$, and the eigenvalues are plotted as
functions of the angular half width of the nodal pocket centered on
$\theta = 0$.  The vertical line shows the nodal surface of the exact
excited state $\Psi_2$. }
\label{fig4}
\end{figure}

\begin{figure}
\begin{center}
\leavevmode
\epsfxsize=8cm \epsfbox{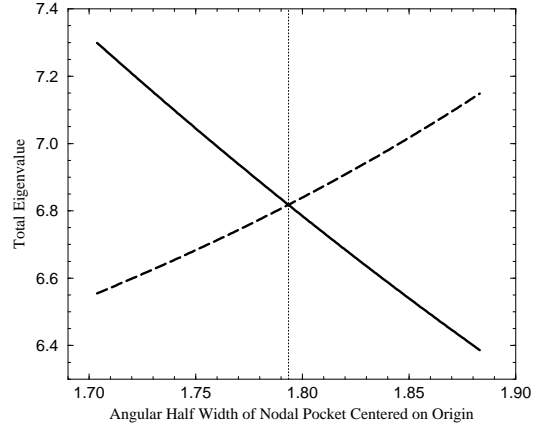}
\end{center}
\caption{ The total pocket eigenvalues $E$ of the two nodal pockets of
the separable example from Sec.~\ref{sec:sepexample}.  The Hamiltonian
is that of Eq.~(\ref{eq:sepham}) with $V({\bf
r})=\cos(3\theta)/\rho^2$, and the eigenvalues are plotted as
functions of the angular half width of the nodal pocket centered on
$\theta = 0$.  The vertical line shows the nodal surface of the exact
excited state $\Psi_2$. }
\label{fig5}
\end{figure}

This analysis is confirmed by the results of a full calculation given
in Fig.~(\ref{fig4}), which shows how the angular pocket eigenvalues
$E_{\theta}$ of the two pockets depend on the angular half width of
the nodal pocket centered on $\theta = 0$.  As expected, $E_{\theta}$
is too large when the nodal pocket is too small and vice-versa.
Fig.~(\ref{fig5}) shows the dependence of the total pocket eigenvalues
$E$ on the angular half width of the nodal pocket centered on $\theta
= 0$, confirming that an increase in angular half width gives rise to
a decrease in total eigenvalue and vice-versa.  The numerical results
therefore support the conclusions of the variational argument.

Note, finally, that Eq.~(\ref{eq:thetaeq}) can be interpreted as the
Schr\"{o}dinger equation for a one-dimensional crystal with periodic
boundary conditions, in which case $\Psi_1$ and $\Psi_2$ are real
linear combinations of Bloch waves with equal and opposite crystal
momenta.  Seen from this viewpoint, the degeneracy of $\Psi_1$ and
$\Psi_2$ arises from the time-reversal (complex-conjugation) symmetry
of the real Hamiltonian rather than from its spatial symmetry, but the
failure of the symmetry-constrained variational principle is still
apparent.  This confirms the assertion made in Sec.~\ref{sec:d>1}: a
DMC simulation using a real trial state constructed from Bloch states
with equal and opposite crystal momenta may yield an energy below that
of the lowest exact eigenstate with that crystal momentum.

\section{Conclusions}
\label{sec:Conclusions}

The main lesson to be learned from this work is that symmetry
arguments cannot be applied to fixed-node DMC unless the symmetries of
both the Hamiltonian and the nodal surface of the trial wave function
are taken into account.  The fixed-node DMC algorithm solves the
Hamiltonian subject to the boundary conditions imposed by the trial
nodal surface, and so the relevant symmetry group ${\mathcal G}_{{\rm
FN}}$ contains only those symmetry operations that leave both the
Hamiltonian and the trial nodal surface (boundary conditions)
invariant.  This ``group of the fixed-node Hamiltonian'' is a subgroup
of the more familiar ``group of the Hamiltonian'', ${\mathcal G}$.

If the trial function transforms according to a real one-dimensional
irreducible representation $\Gamma^r$ of ${\mathcal G}$, all symmetry 
operations in ${\mathcal G}$ simply multiply the trial function by a 
real number.  This does not change the nodal surface and hence 
${\mathcal G} = {\mathcal G}_{{\rm FN}}$.  The symmetry-constrained 
variational principle then implies: (i) that the fixed-node DMC energy 
is greater than or equal to the eigenvalue of the lowest energy exact 
eigenstate that transforms according to the same one-dimensional 
irreducible representation as the trial state; and (ii) that the error 
in the DMC energy is in general second order in the difference between 
the nodal surfaces of the lowest energy exact eigenstate that transforms
as $\Gamma^r$ and the trial function.

If the irreducible representation to which the trial function belongs
is of dimension greater than one, it is inevitable that some of the
symmetry operations from ${\mathcal G}$ will change the trial nodal
surface.  The nodal surfaces of the states $\Psi_1$ and $\Psi_2$ shown
in Fig.~(\ref{fig1}) are examples of this.  The symmetry-constrained
variational principle need not apply in such cases, because the
symmetrized state $\bar{\Psi}_{\alpha}^{r}$ used in its derivation
need not be zero everywhere on the trial nodal surface.  The delta
functions produced when the kinetic energy operator is applied to the
fixed-node pocket ground state $\phi_{\alpha}$ from
Eq.~(\ref{eq:eigensoln}) may therefore contribute to the expectation
value in Eq.~(\ref{eq:Ealphap}).

In such cases, a weaker version of the symmetry-constrained
variational principle can be obtained by re-analyzing the problem
using only the symmetries in the subgroup ${\mathcal G}_{\rm FN}$.
The idea is to forget all the symmetry operations which change the
nodes of the trial function, and consider only those which leave the
trial nodal surface invariant.  The symmetry-constrained variational
principle then applies as long as the symmetries are labeled using the
irreducible representations of ${\mathcal G}_{\rm FN}$.

In summary, a useful DMC variational principle exists whenever the
trial state transforms according to a one-dimensional irreducible
representation $\Gamma^r$ of ${\mathcal G}$ or any subgroup of
${\mathcal G}$. In many DMC simulations, the nodal surface of the
trial state is the same as that of an eigenstate of an approximate
Hamiltonian such as the local-density-functional Hamiltonian.  If the
state used to define the nodal surface is the lowest energy eigenstate
with symmetry $\Gamma^r$ of an approximate Hamiltonian with a
reasonable local potential, the generalized tiling theorem discussed
in Sec.~\ref{sec:tiling} shows that all the nodal pockets are
equivalent by symmetry.  The DMC energy is therefore independent of
the initial distribution of walkers among the nodal pockets.

The ordinary fixed-node approximation provides a good example of these
ideas.  The many-electron ground state is never the overall ground
state (which is bosonic), and may be degenerate, in which case we
cannot prove the existence of a variational principle by analyzing the
problem using the full symmetry group.  We can, however, use the
permutation group, which is always a subgroup of the full symmetry
group.  The many-electron trial function transforms according to the
one-dimensional antisymmetric irreducible representation of this
subgroup.  The weaker variational principle therefore guarantees that
the fixed-node DMC energy is greater than or equal to the energy of
the many-electron ground state; and the generalized tiling theorem
guarantees that the exact many-electron ground state possesses the
tiling property with respect to permutations.  This shows that the
ground-state versions of the fixed-node variational
principle~\cite{reynolds} and tiling theorem~\cite{tiling} may be
regarded as special cases of the more general versions discussed in
this paper.

The different members of a set of trial functions forming a basis for
a multi-dimensional irreducible representation of ${\mathcal G}$ have
different nodal surfaces and need not all produce the same fixed-node
DMC energy.  The strength of the weaker variational principle may
therefore be optimized by using specific linear combinations of these
basis functions.  The best linear combinations transform according to
one-dimensional irreducible representations of subgroups of ${\mathcal
G}$, and may be found following the procedure explained in
Sec.~\ref{sec:d>1}.  If this procedure is not carried out, trial
functions belonging to multi-dimensional irreducible representations
of ${\mathcal G}$ usually have nodal surfaces with very little spatial
symmetry.  In many cases, the only symmetry operations that leave the
nodal surface invariant are the elements of the permutation group, and
the only variational principle that survives is the one relating to
the many-electron ground state.

\section{Acknowledgements}
\label{sec:Acknowledgements}

We thank the Engineering and Physical Sciences Research Council, UK,
for financial support under grants GR/L40113 and GR/M05348.

\newpage

\begin{references}
\bibitem{lie} The results presented in this section also hold for
infinite order compact Lie groups, and hence are sufficiently general
to encompass most fixed-node DMC simulations of real electronic
systems.
\end{references}
\end{document}